\documentclass[sigconf, screen]{acmart}

\setcopyright{acmlicensed}
\copyrightyear{2026}
\acmYear{2026}
\acmDOI{XXXXXXX.XXXXXXX}
\acmConference[KDD~Workshop'26]{KDD 2026 Workshop on AI Data Scientist}{August 2026}{Toronto, Canada}
\acmISBN{978-1-4503-XXXX-X/2026/08}

\usepackage[ruled,lined,linesnumbered]{algorithm2e}
\usepackage{colortbl}
\usepackage{arydshln}
\usepackage{fontawesome5}
\usepackage{enumitem}
\usepackage{tabularx}
\usepackage{ragged2e}
\usepackage{multirow}
\usepackage{subcaption}

\usepackage{xspace}
\newcommand{\system}{\textsc{sketch-plot}\xspace}

\definecolor{bestblue}{RGB}{220, 235, 250}
\definecolor{secondgreen}{RGB}{235, 245, 230}
\definecolor{bestyellow}{RGB}{255, 243, 205}

\usepackage{pifont}

\title{\system: Progressive Editing for Text-to-Image Academic Figures}

\author{Yinghao Tang}
\affiliation{%
  \institution{State Key Lab of CAD\&CG, Zhejiang University}
  \city{Hangzhou}
  \country{China}}
\email{yinghaotang@zju.edu.cn}

\author{Yupeng Xie}
\affiliation{%
  \institution{HKUST(GZ)}
  \city{Guangzhou}
  \country{China}}
\email{yxie740@connect.hkust-gz.edu.cn}

\author{Yingchaojie Feng}
\affiliation{%
  \institution{National University of Singapore}
  \city{Singapore}
  \country{Singapore}}
\email{feng.y@nus.edu.sg}

\author{Tingfeng Lan}
\affiliation{%
  \institution{University of Virginia}
  \city{Charlottesville}
  \state{VA}
  \country{USA}}
\email{tingfeng@virginia.edu}

\author{Jiale Lao}
\affiliation{%
  \institution{Cornell University}
  \city{Ithaca}
  \state{NY}
  \country{USA}}
\email{jiale@cs.cornell.edu}

\author{Wei Chen}
\affiliation{%
  \institution{State Key Lab of CAD\&CG, Zhejiang University}
  \city{Hangzhou}
  \country{China}}
\email{chenvis@zju.edu.cn}

\renewcommand{\shortauthors}{Tang et al.}

\begin{document}

\begin{abstract}
Text to image (T2I) models such as \texttt{gpt-image-2} can now generate publication grade academic figures from a short prompt, but the output is a flat raster: a user who wants to change one arrow, one label, or one icon has to regenerate the whole image, which also disturbs the parts they wanted to keep. We present \system, an interactive system that closes this controllability gap with a three layer progressive editing pipeline: a generated PNG, an addressable puzzle of editable pieces, and a per piece SVG. The user stops at the layer that gives them enough control for the change at hand, so the cost of decomposition and vectorisation is paid only on the pieces that need it. Realising this pipeline is not trivial. General segmentation models lack the semantic discriminability to decompose a research figure cleanly, and end to end image vectorisation produces incomplete shapes and loses semantic structure. We therefore route both stages through a \textbf{human in the loop} interface that lets the user accept, refine, or reject decomposition and vectorisation decisions on a piece by piece basis. We validate the design with an expert user study, in which participants found \system effective for making targeted edits to AI generated academic figures and preferred it over regenerating the whole image. A demonstration video is available at \textbf{\url{https://paper-plot.dev/sketch}}.
\end{abstract}



\maketitle
\section{Introduction}

\begin{figure}[t]
    \centering
    \includegraphics[width=\linewidth]{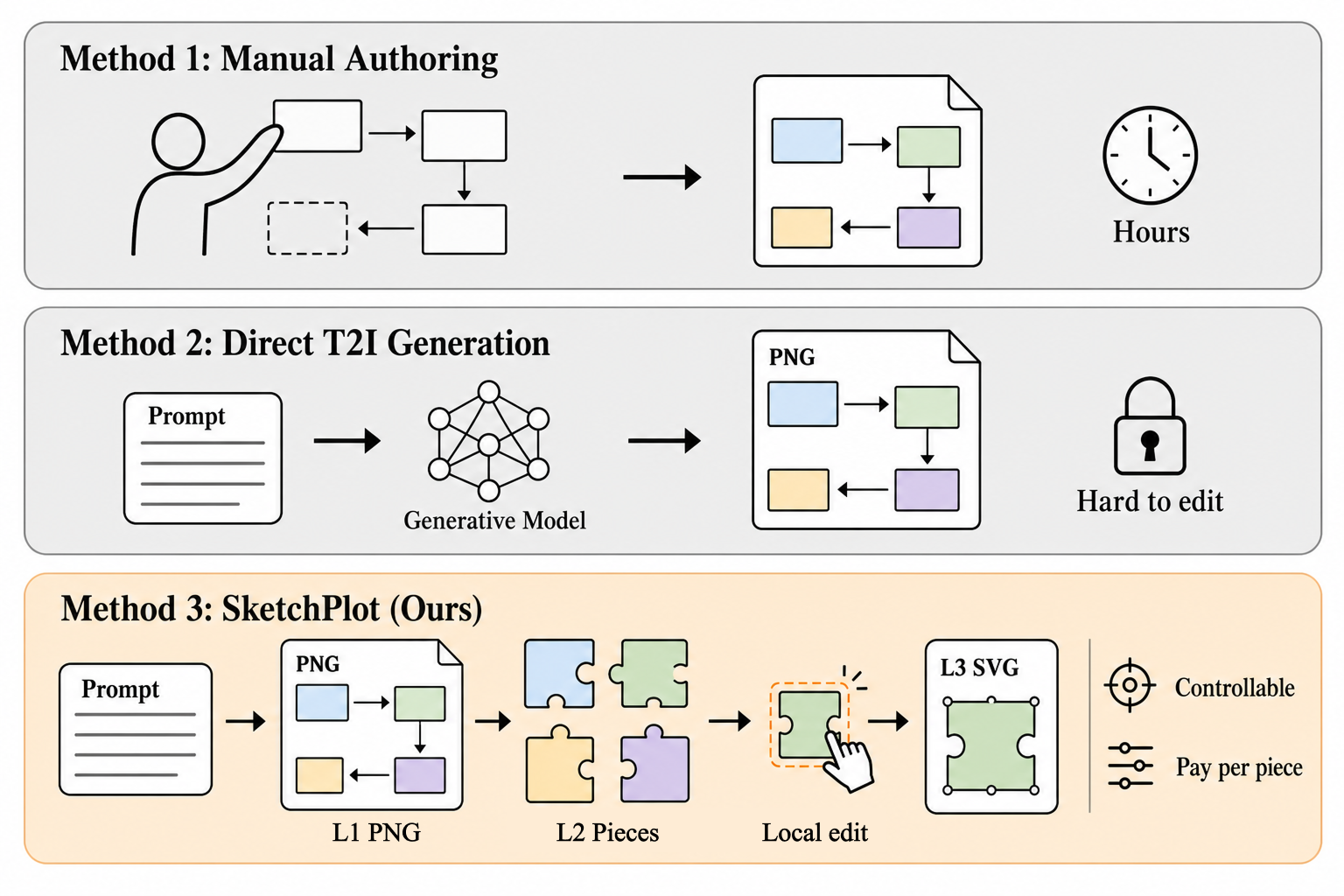}
    \caption{Three approaches to creating editable academic figures: manual authoring, direct text-to-image generation, and human-guided progressive editing through \system.}
    \Description{A three-row comparison diagram contrasting manual authoring, direct text-to-image generation, and SketchPlot's progressive PNG, pieces, and SVG workflow.}
    \label{fig:method-comparison}
\end{figure}

Academic figures such as flowcharts, pipeline diagrams, and architecture overviews are essential for communicating complex ideas in scholarly papers~\cite{lee2017visstyle,franconeri2021science,tang2026vividoc,chen2025chartmark}. However, creating these figures remains time consuming, often requiring hours of manual layout in tools such as PowerPoint, Keynote, or TikZ~\cite{belouadi2023automatikz}.

A promising approach is to delegate this work to text to image (T2I) models. Modern T2I models such as \texttt{gpt-image-2}~\cite{openai_gptimage}, \texttt{qwen-image}~\cite{qwen_image}, and \texttt{gemini-2.5-flash-image}~\cite{gemini_image} can produce visually polished figures from a short text prompt in seconds, and recent benchmarks confirm that the strongest among them reach publication grade quality on a substantial fraction of cases~\cite{tang2026igenbench,li2026deepeye}. As Figure~\ref{fig:method-comparison} illustrates, this creates a tradeoff between slow but editable manual authoring and fast but hard to control generation. The fundamental limitation is that the generated output is a flat raster with no exposed structure. As a consequence, even a minor correction, such as fixing a label or changing an arrow direction, requires regenerating the entire figure, which also alters the parts that were already correct. The result is a polished figure that is not editable, a property we refer to as the \emph{controllability gap} of T2I generation.

In this paper, we aim to solve this gap with a three layer progressive controllability pipeline, shown in the bottom row of Figure~\ref{fig:method-comparison}. The first layer is the raster PNG returned by the T2I model. The second layer is a \emph{puzzle}: a complete partition of the canvas into addressable pieces that can be hovered, selected, resized along shared edges, copied, and re prompted, while the rest of the image stays fixed at the pixel level. The third layer is an SVG view in which a piece's text, colours, and strokes become attributes that the user can edit directly, without invoking any image model. The user stops at the layer that gives them enough control for the change at hand, so the cost of decomposition and vectorisation is paid only on the pieces that need it.

Realising this pipeline is not trivial. First, general segmentation backbones lack the semantic discriminability to recover panels, header bars, and labelled boxes from a research figure as coherent units~\cite{espinosa2024samantics}. Second, end to end image vectorisation, applied to a whole figure, produces incomplete shapes and loses semantic structure~\cite{bing2024deepicon}. We address both with a \textbf{human in the loop} interface that sits between the user and the underlying models. For the first challenge, instead of asking a segmentation model to produce a perfect decomposition, we ask a vision language model for a coarse layout and let the user refine it directly in the puzzle interface, fixing any decomposition mistake before moving on. For the second, instead of vectorising the whole figure in one shot, we expose vectorisation as a per piece operation invoked only on the pieces the user selects. Together, these mechanisms give the user fine grained control over what gets edited and at what fidelity, while keeping any model error local to a single piece rather than the whole figure.

We summarise our contributions as follows:

\begin{itemize}[leftmargin=*,itemsep=2pt,topsep=2pt]
    \item We propose a three layer progressive controllability pipeline (PNG, puzzle, SVG) for post hoc editing of T2I generated academic figures, with a human in the loop design that sidesteps the failure modes of end to end segmentation and vectorisation.
    \item We present \system, an open source web based instantiation of the pipeline that integrates a T2I model, a vision language model for layout, and a per piece vectoriser into a single editing interface.
    \item We validate the design with an expert user study, in which participants found \system effective for making targeted edits to AI generated academic figures and preferred it over regenerating the whole image.
\end{itemize}

A demonstration video is available at \textbf{\url{https://paper-plot.dev/sketch}}.

\section{Related Work}
\label{sec:related}

\noindent
\textbf{Text-to-Image Figure Generation.}
Recent text-to-image (T2I) models~\cite{openai_gptimage,gemini_image,rombach2022high}, such as gpt-image-2~\cite{openai_gptimage} and gemini-2.5-flash-image~\cite{gemini_image}, can produce visually polished academic figures from a short prompt in seconds, and dedicated reliability benchmarks have begun to assess them, e.g., IGenBench~\cite{tang2026igenbench}. Their output, however, is a flat raster with no exposed structure: individual elements cannot be addressed, and correcting even a single label requires regenerating the whole figure, which alters parts the author wished to preserve. Code-driven alternatives~\cite{belouadi2023automatikz}, such as AutomaTikZ~\cite{belouadi2023automatikz}, restore full editability by generating TikZ source, but the visual style is bound by what the underlying language can express and the authoring curve remains steep. \system sits at the opposite end of this trade-off: visual quality comes from the T2I model, and structure is recovered afterwards.

\smallskip
\noindent
\textbf{Image Editing and Segmentation.}
Mask-conditioned image editing methods~\cite{brooks2023instructpix2pix,openai_gptimage}, such as InstructPix2Pix~\cite{brooks2023instructpix2pix}, allow a user to modify a region of an image given a mask and a text instruction, but the mask itself must be painted by hand without any knowledge of the figure's semantic structure. General-purpose segmentation models~\cite{kirillov2023sam,sam2}, such as SAM~\cite{kirillov2023sam}, produce accurate pixel-level masks yet over-segment text into individual glyphs, miss thin arrows below their area threshold, and emit no semantic label for what a region means. \system supplies the missing layer: a VLM-derived, semantically named partition of the canvas that lets the user address a region by its semantic role rather than by a brush stroke. This named partition directly enables region-level inpainting: the user selects a region by name and issues an edit prompt, and its bounding box automatically serves as the mask.

\section{System}
\label{sec:system}

\begin{figure*}[t]
  \centering
  \includegraphics[width=\textwidth]{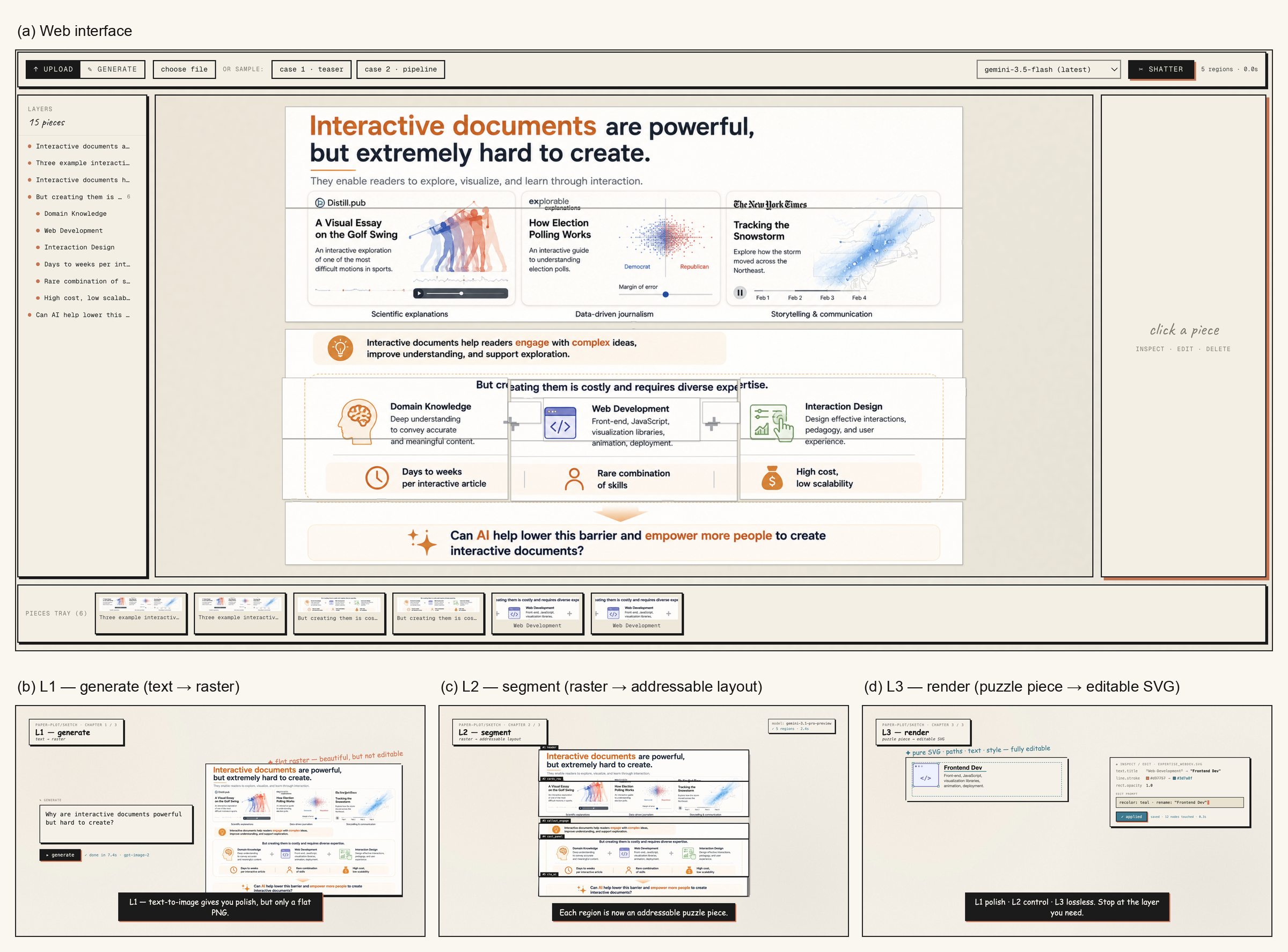}
  \caption{System overview of \system. (a) The web interface exposes the
    three layers through a single canvas: a top bar drives generation
    and the \textsc{shatter} action; the centre canvas hosts the raster
    figure; the left \emph{layers} panel and bottom \emph{pieces tray}
    expose the segmented hierarchy; the right inspector hosts per-piece
    \emph{inspect / edit / delete} actions. (b, c, d) The three
    progressive editing stages. \textbf{L1 (generate)} turns a text
    prompt into a flat raster figure. \textbf{L2 (segment)} asks a VLM
    to propose regions and uses a residual-completion pass to turn the
    raster into a canvas of addressable \emph{puzzle pieces}.
    \textbf{L3 (render)} vectorises a selected piece on demand,
    exposing text, stroke, and style as first-class handles. Users
    stop at whichever layer their task demands.}
  \label{fig:system}
\end{figure*}

\subsection{Overview}
\label{sec:overview}

\system is organised as a three-layer pipeline driven by a single
human-in-the-loop interface (Figure~\ref{fig:system}). Given a short
text prompt, layer~\textbf{L1} calls a text-to-image model and returns
a flat raster figure. Our current implementation routes through
OpenRouter and uses \texttt{gpt-image-2} at $1536 \times 1024$, but the
backend is interchangeable and we have also 
\texttt{gemini-3.1-flash-image-preview}. The raster looks finished
but is opaque. Every box, arrow, and label is fused into a single PNG,
so any subsequent edit forces the user back to the prompt. Layer
\textbf{L2} closes that gap. A vision-language model (default
\texttt{gemini-3.5-flash}) is asked to read the raster and emit a
coarse layout: a small set of three to ten top-level bounding boxes
with semantic labels (e.g.\ \texttt{header}, \texttt{cards\_row},
\texttt{cost\_panel}). Layer~\textbf{L3} is invoked on demand. For any
selected piece, the system runs an image-to-SVG pass that converts the
local raster region into editable vector primitives, so text, stroke,
fill, and geometry become first-class handles instead of pixels.

The VLM output by itself is not yet a controllable surface. The
returned boxes have gaps, occasional overlaps, and ignore everything
the model judged unimportant, so they cannot be used as a partition.
We close the layout into a \emph{puzzle} with a deterministic
residual-completion step. Named regions are sorted by area in
ascending order so that smaller, more specific boxes win any overlap
with their larger ancestors. We then walk the regions in that order
and clip each one to the still-free area of the canvas using
axis-aligned rectangle subtraction, which keeps the largest contiguous
fragment as the visible piece for that label and re-emits any leftover
slivers as background pieces. Whatever free area remains at the end is
sliced into background rectangles as well. The output is an exact
tiling of the canvas: every pixel belongs to exactly one piece, every
piece carries either a semantic label or a \texttt{background} tag,
and every piece exposes the same interaction vocabulary. This
invariant is what lets the rest of the system treat ``imperfect VLM
output'' as a feature rather than a blocker, since the user can
absorb, split, or re-label background pieces without ever leaving the
canvas.

\subsection{User Interface}
\label{sec:ui}

The interface (Figure~\ref{fig:system}a) is a single-page canvas
organised into four regions. The \textbf{top bar} hosts the three
pipeline controls: \textsc{upload} an existing figure or
\textsc{generate} one from a prompt, pick a backbone model, and
trigger \textsc{shatter} (the L2 segmentation pass). The
\textbf{centre canvas} always shows the current raster, with each
puzzle piece drawn as a translucent overlay so that hovering reveals
the piece boundary and clicking selects it. The \textbf{layers panel}
on the left and the \textbf{pieces tray} at the bottom both expose the
same L2 hierarchy: the panel as a textual outline (for navigation by
label), the tray as thumbnails (for navigation by appearance). The
\textbf{inspector} on the right is the editing surface for the
currently selected piece.

The inspector exposes four actions on a piece, ordered by increasing
commitment. \emph{Inspect} reveals the piece's metadata (label, type,
bounding box, parent) and a thumbnail crop. \emph{Drag-edge} lets the
user nudge the boundary shared with a neighbour. The drag is resolved
by re-running \textsc{tilify} with the dragged region pinned to its
new bounds, so neighbours give way and the canvas is re-tiled in a
single pass without leaving holes. \emph{Re-prompt} sends the piece's
crop, mask, and a short instruction to an image-edit model and pastes
the result back into that region only, leaving every other piece
untouched at the pixel level. \emph{Vectorise} invokes L3 and replaces
the piece's raster with editable SVG. Because all four regions stay
live at once, the user can move between layers (regenerate L1 from a
revised prompt, re-shatter L2 with a different backbone, vectorise an
L3 piece) without leaving the page or losing the rest of the canvas.

\section{User Study}

We conducted a user study to evaluate whether \system helps users create and refine AI-generated academic figures in a controllable way across its L1--L3 workflow.

\paragraph{Participants.}
We recruited three participants from artificial intelligence and human-computer interaction. All participants had created figures for academic papers or presentations, and none had used \system before the study.

\paragraph{Procedure.}
Each session lasted approximately 40 minutes. We first gave participants a brief introduction to \system and demonstrated the basic interaction flow. Participants then completed the figure authoring task while thinking aloud. After the task, they answered five questions on a 5-point Likert scale, covering ease of learning, ease of use, the L1--L3 workflow, the L2 decomposition stage, and the L3 rendering stage, and joined a short semi-structured interview about their experience.

\paragraph{Tasks.}
Participants were asked to create an academic figure based on their own research work using \system. The target figure could be a method overview, pipeline diagram, system architecture figure, or conceptual framework. Participants first described the figure they wanted to create, then used \system to generate an initial figure and refine selected parts through the L1--L3 workflow.

\begin{figure}[t]
    \centering
    \includegraphics[width=\linewidth]{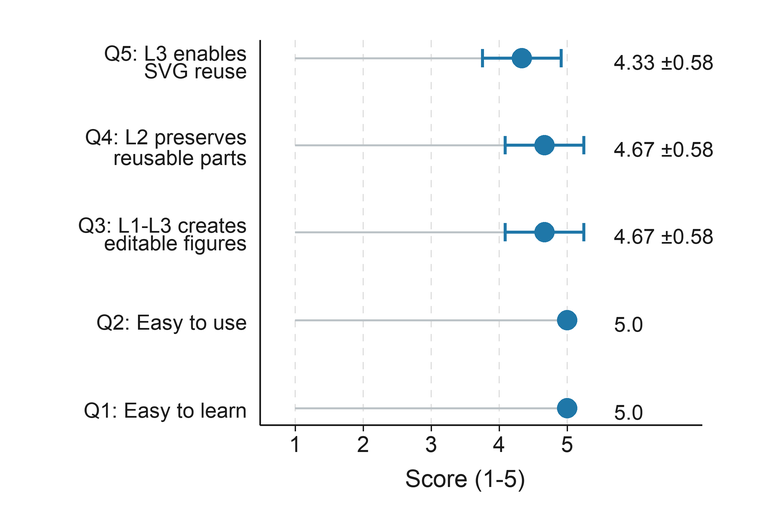}
    \caption{Participant ratings of \system on five study questions. Values show mean and standard deviation.}
    \Description{A horizontal dot-and-error-bar chart showing participant ratings of \system on five questions from 1 to 5.}
    \label{fig:user-study}
\end{figure}

\paragraph{Results and Feedback.}
Overall, participants responded positively to \system. As shown in Figure~\ref{fig:user-study}, participants rated the system highly on ease of learning and ease of use, with both questions receiving a mean score of 5.0. They also agreed that the L1--L3 workflow helped them turn a research idea into an editable academic figure ($4.67 \pm 0.58$), and that the L2 decomposition stage helped preserve and reuse satisfactory parts of the figure ($4.67 \pm 0.58$). Participants gave a slightly lower but still positive rating to the L3 rendering stage for converting selected raster pieces into SVG elements for further editing and reuse ($4.33 \pm 0.58$). In the interviews, participants appreciated that useful parts of a generated figure could be kept as reusable pieces instead of being lost during regeneration. They also found the decomposition helpful for reasoning about the figure as a set of academic visual units, such as panels, labels, icons, and arrows. For L3, participants valued the possibility of moving selected parts into an editable SVG form, while noting that complex icons or dense visual regions sometimes still required additional refinement.

\paragraph{User Suggestions.}
Participants suggested several improvements. They wanted an option to further decompose a selected region for finer control inside a larger panel, more precise boundary snapping for dense, text-heavy layouts, and editing affordances such as undo, edit preview, and version history.


\section{Future Work}
\label{sec:future}

We see \system as a step toward a broader practice for authoring
academic figures in which T2I and VLM models act as collaborators the
author can steer at the level of individual pieces. We hope the
puzzle paradigm extends beyond the flowcharts and architecture
diagrams that drove the current design. Tables, experimental plots,
and code-driven figures share the same controllability gap, but each
calls for a different vocabulary of editable units: rows, headers,
and cell groups for tables; axes, series, and annotations for plots;
and the underlying script alongside the rendered image for
code-driven figures. Working out these per-domain vocabularies and
their associated best practices is the direction we plan to pursue
next.

Two known limitations point to concrete next steps. First, current
VLMs are still imperfect at parsing structured visual content
\cite{espinosa2024samantics}: panels can be fused, sub-panels
dropped, and bounding boxes are sometimes a few pixels short. The
puzzle absorbs these failures as background pieces the user can fix,
and we plan to reduce the cost with prompting refinements and a
pass that snaps box edges to image gradients. Second, image-to-SVG
conversion in general remains weak on fine edge detail such as icons,
glyphs, and thin connectors \cite{bing2024deepicon,belouadi2023automatikz};
closing this gap as the underlying models improve is a direction we
plan to track.

\section{Conclusion}

We proposed the puzzle paradigm for post hoc editing of T2I generated academic figures, defined by two principles: every pixel belongs to exactly one editable piece, and every piece exposes the same interaction vocabulary. The paradigm reframes layout extraction as ``any layout, plus a residual completion'' rather than ``perfect layout'', so imperfect VLM output stops being a blocker for a controllable editing surface. We instantiated the paradigm in \system, a three layer progressive editing pipeline (PNG, puzzle, SVG) wrapped in a human in the loop interface, and validated the design with an expert user study in which participants found \system effective for making targeted edits to AI generated academic figures and preferred it over regenerating the whole image. 

\clearpage
\bibliographystyle{ACM-Reference-Format}
\bibliography{custom}

\newpage
\appendix

\end{document}